\documentclass[12pt]{iopart}
\usepackage{graphicx}
\usepackage{iopams}

\begin{document}

\begin{flushright}
\end{flushright}


\title{Cosmological implications of supersymmetric axion models}
\author{Masahiro Kawasaki, Kazunori Nakayama, Masato Senami
\footnote{Now, at Department of Micro Engineering, Kyoto University.}
}
\address{Institute for Cosmic Ray Research,
University of Tokyo,
Kashiwa 277-8582, Japan}
\date{\today}

\begin{abstract}
We derive general constraints on supersymmetric extension of axion
models, in particular paying careful attention to the cosmological
effects of saxion.  It is found that for every mass range of the
saxion from keV to TeV, severe constraints on the energy density of the
saxion are imposed. Together with constraints from axino we obtain
stringent upper bounds on the reheating temperature.
\end{abstract}

\maketitle

\setcounter{footnote}{0}

\section{Introduction}

One of the main problems of the standard model is the strong violation
of CP invariance due to non-perturbative effects of quantum
chromodynamics (QCD).  In general, QCD effects generates the term in the
lagrangian such as
\begin{equation}
   \mathcal L_\theta = \frac{\theta g_s^2}{32\pi^2} 
    G_{\mu \nu}^a \tilde G^{\mu \nu a},
\end{equation}
where $g_s$ is the QCD gauge coupling constant, $G_{\mu \nu}^a$ is the
field strength of the gluon, and $\tilde G^{\mu \nu a}=\epsilon ^{\mu
\nu \rho \sigma}G_{\rho \sigma}^a/2 $.  Experimentally $\theta$ must be
smaller than about $10^{-9}$, but in the standard model there seems to
be no theoretical reasons that $\theta$ must be so small.  This is the
well-known strong CP problem.  As a solution to the strong CP problem,
Peccei and Quinn \cite{Peccei:1977hh} introduced anomalous global U(1)
symmetry, which we denote U(1)$_{\rm PQ}$.  When this U(1)$_{\rm PQ}$ is
broken spontaneously, there appears a pseudo-Nambu Goldstone boson
called axion \cite{Kim:1986ax}.  The axion dynamically cancels the
$\theta$-term effectively, and no strong CP violation is observed in
true vacuum.  In order not to contradict with terrestrial experiments,
astrophysical \cite{Raffelt:1990yz} and cosmological arguments
\cite{Preskill:1982cy}, the breaking scale of PQ symmetry $F_a$ should
lie in the range $10^{10}~{\rm GeV} \lesssim F_a \lesssim 10^{12}~{\rm
GeV}$.  If $F_a$ is close to this upper bound, the axion is an
interesting candidate for the cold dark matter.

On the other hand, another problem in the standard model is the
quadratic divergence of the radiative correction to the Higgs mass.  In
the standard model, the mass of the Higgs is not protected by any
symmetry, and hence naturally the Higgs boson is expected to obtain the
mass of the cut-off scale.  Thus to obtain hierarchically small mass
scale down to the weak scale requires unnatural fine-tuning.
Supersymmetry (SUSY) \cite{Nilles:1983ge} is the most motivated solution
to this problem, since SUSY protects the mass of scalar fields and the
weak scale becomes stable against radiative correction.  From cosmological
points of view, SUSY also provides interesting candidates for the dark
matter.  Due to the $R$-parity conservation, under which standard model
particles have the charge $+1$ and their superpartners have $-1$, the
lightest SUSY particle (LSP) is stable.  So if the LSP is
neutralino or gravitino, they are the dark matter candidates.

Therefore, it seems reasonable to combine these
two paradigms.  In fact, axion models are easily extended to
implement SUSY.  As we will see, in SUSY extensions of the
axion models, many non-trivial cosmological consequences arise.  In SUSY
axion models, both the scalar partner of the axion, {\it saxion}, and
the fermionic superpartner of the axion, {\it axino}, have significant
effects on cosmology.  But there have not been many studies which treat
both of them in spite of their importance
\cite{Kim:1992eu,Lyth:1993zw,Hashimoto:1998ua,Asaka:1998xa}. Moreover, these earlier
works were based on specific models and only the restricted parameter
regions were investigated.  In this paper, we investigate all possible
mass range of the saxion and corresponding various cosmological bounds.
Our results are easily applied to any axion models with slight
modifications, and hence provide general cosmological constraints.
Furthermore, since the saxion and axino densities depend on the
reheating temperature $T_R$ after inflation, we can obtain constraints
on $T_R$.  The important result is that the upper bound on $T_R$ becomes
more stringent due to the late-decaying saxion and axino,
than that from the usual gravitino problem.

This paper is organized as follows.  In Sec.~\ref{sec:axionmodel} we
explain the dynamics of the saxion and its properties.  In
Sec.~\ref{sec:constraints} we derive various cosmological constraints on
SUSY axion models, especially upper bound on the reheating
temperature.  Implications for the dark matter in the SUSY axion models is
discussed in Sec.~\ref{sec:others}.
In Sec.~\ref{sec:light} we briefly comment on the ultra-light gravitino scenario,
where the gravitino is $O(10)$~eV.
We conclude in Sec.~\ref{sec:conclusion}.

\section{Supersymmetric axion models}  
\label{sec:axionmodel}

The axion is the pseudo-Nambu-Goldstone boson which appears due to the
spontaneous breaking of PQ symmetry.  The axion obtains a mass from the
effect of quantum anomaly, but this contribution is very small.  It is
estimated as $m_{a0} \sim 6\times10^{-6}~{\rm eV}(10^{12}~{\rm GeV} /
F_a )$.

In SUSY extensions of axion models, the axion field forms supermultiplet,
which contains a scalar partner (saxion) and fermionic superpartner
(axino) of the axion \cite{Rajagopal:1990yx}.  The saxion mass is
expected to be of the order of the gravitino mass ($m_{3/2}$).  On the
other hand the axino mass somewhat depends on models, but generically can
be as large as $m_{3/2}$.  Although the interactions of both particles
with standard model particles are suppressed by the PQ scale $F_a$, they
may cause significant effects on cosmology as will be seen in
Sec.~\ref{sec:constraints}.

\subsection{Dynamics of the saxion }  
\label{sec:dynamics} 

In SUSY models, superpotentials must satisfy the holomorphy.  When the real
U(1)$_{{\rm PQ}}$ symmetry is combined with the holomorphic property of
the superpotential, it is extended to complex U(1) symmetry, which
inevitably includes scale transformation \cite{Kugo:1983ma}.  The
invariance under the scale transformation means the existence of a flat
direction along which the scalar field does not feel the scalar potential.  The
saxion corresponds to such a flat direction of the potential.
But SUSY breaking effects lift the flat direction and the saxion
receives a mass of order $m_{3/2}$.\footnote{
In gauge-mediated SUSY breaking models, the saxion receives a 
logarithmic potential from gauge-mediation effects. 
If the PQ scalar is stabilized by the balance between the logarithmic potential and
the gravity-mediation effect, $m_s \sim m_{3/2}$ still holds
\cite{Asaka:1998ns} .
}
The saxion field can develop to large field value during inflation, and
begins to oscillate around its minimum when the Hubble parameter $H$
becomes comparable to the saxion mass $m_s \sim m_{3/2}$.  In general
such coherent oscillation of the saxion field has large energy density
and hence its late decay may have significant effects on cosmology
\cite{Kim:1992eu,Hashimoto:1998ua,Asaka:1998xa,Banks:2002sd}.

As an example, let us consider a model with the following
superpotential:
\begin{equation}
	W = \lambda X (\Phi \bar{\Phi}-F_a^2) ,  \label{Wsaxion}
\end{equation}
where the superfields $X,\Phi$ and $\bar{\Phi}$ have the PQ charges
$0,+1,-1$ respectively.  The saxion direction is easily identified as
$\Phi \bar{\Phi}=F_a^2$ with $X=0$.\footnote{
If the $A$-term contribution $V_A \sim m_{3/2} \lambda X F_a^2+{\rm h.c.}$ is included, the $X$ field can have the VEV
$|X|\sim m_{3/2}$.  But this does not modify the following arguments. }
Including SUSY breaking mass terms due to gravity-mediation effects, the
scalar potential can be written as
\begin{eqnarray}
   V = &m_{3/2}^2\left ( c_X|X|^2 +c_1|\Phi|^2+c_2|\bar{\Phi}|^2 \right )  
    \nonumber \\
       &+|\lambda|^2 \left \{ |\Phi \bar{\Phi}-F_a^2 |^2 
        +|X|^2\left ( |\Phi|^2+|\bar{\Phi}|^2  \right ) \right \},   
	\label{Vsaxion}
\end{eqnarray}
where $c_X, c_1$ and $c_2$ are $O(1)$ constants which are assumed to be
positive.  The potential minimum appears at $|\Phi| \sim |\bar{\Phi}|
\sim F_a$, but the initial amplitude of the saxion remains undetermined.
In general, if the saxion remains light during inflation, the field
value may naturally take the value of the order of the reduced Planck scale
$M_P$.  
But the initial amplitude can be suppressed by introducing the
Hubble-induced mass terms, which are induced by the saxion coupling
with inflaton field through supergravity effect, given by
\begin{equation}
   V_H = H^2\left ( c_X^{\prime}|X|^2 +c'_1|\Phi|^2+c'_2|\bar{\Phi}|^2 \right )
\end{equation}
where $c_X^{\prime}, c'_1$ and $c'_2$ are $O(1)$ constants.  If either $c'_1$
or $c'_2$ are negative, the $\Phi$ or $\bar{\Phi}$ field roll away to
$\sim M_P$ during inflation.  
But if both coefficients are positive, the
potential minimum during inflation is also given by $|\Phi| \sim
|\bar{\Phi}| \sim F_a$, which almost coincides with the low energy true
minimum.  Since there is a priori no reason that we expect that
$c_1/c_2$ is exactly equals to $c'_1/c'_2$, these two minima are separated
by
\begin{equation}
   s \sim \left [\left ( \frac{c_1}{c_2} \right )^{1/4} 
	   - \left ( \frac{c'_1}{c'_2} \right )^{1/4} \right ] F_a,
\end{equation}
where $s$ denotes saxion field (here we have assumed $c_1>c_2$ and
$c'_1>c'_2$).  This simple model provides one realization of the scenario
for the initial saxion amplitude $s_i$ to be $\sim F_a$.
For the case of $s_i \sim M_P$, the cosmological constraints become much more stringent than the
case of $s_i \sim F_a$, and hence hereafter we consider only the latter case.

For this scenario to work, the Hubble parameter during inflation $H_I$ should be
smaller than $\sim F_a$, since otherwise the large Hubble mass term
during inflation takes the all fields to the origin.  Once they are
trapped at the origin, the saxion has unsuppressed interaction with
particles in thermal bath and get a large thermal mass, which results in
further trap of of the saxion at the origin. Thus, the oscillation epoch
is significantly delayed~\cite{Asaka:1998xa}.\footnote{
This is the case in the KSVZ (or hadronic axion) model~\cite{Kim:1979if}.
In the DFSZ model~\cite{Zhitnitsky:1980tq} such a thermal mass may not arise
because of the small coupling of the PQ scalar, but there arises another
difficulty from domain wall formation.  } 
Because delayed oscillation only makes the cosmological saxion problems
worse, we do not consider such a case.  On the other hand, the axion
field has isocurvature perturbations with amplitude $\sim H_I/(\pi s_i)$
during inflation~\cite{Seckel:1985tj,Linde:1991km}, which leads to a constraint on
$H_I$ as
\begin{equation}
	H_I \lesssim 2\times 10^7~{\rm GeV} ~\theta_i ^{-1}
	\left ( \frac{\Omega_m h^2}{0.13} \right )
	\left ( \frac{s_i}{F_a} \right )
	\left ( \frac{F_a}{10^{12}~{\rm GeV}} \right )^{-0.175},
\end{equation}
where $\theta_i $ denotes the initial misalignment angle of the axion,
$\Omega_m$ denotes the density parameter of the nonrelativistic matter
and $h$ is the present Hubble parameter in units of 100~km/sec/Mpc.
Here we have used the observational constraint that the ratio of the
isocurvature perturbation to the adiabatic one  should be less
than about $0.3$~\cite{Bean:2006qz}.  Thus, as long as we stick to $s_i
\sim F_a$, the requirement $H_I \lesssim F_a$ is in fact valid from
cosmological point of view.\footnote{
The saxion does not give rise to isocurvature fluctuation because it
has large Hubble mass and its quantum fluctuation is suppressed during
inflation. }

Although we have presented a specific model above, the dynamics of the
saxion does not depend on axion models much, once the initial amplitude
and the mass of the saxion are fixed.  The saxion field starts
oscillation at $H\sim m_s$ with initial amplitude $s_i$.  As explained
above, $m_s$ is likely of the order of $m_{3/2}$ and the natural
expectation of the initial amplitude is $s_i \sim F_a$ or $s_i \sim
M_P$.

Now let us estimate the saxion abundance.  First we consider the saxion
abundance in the form of coherent oscillation.  It is independent of the
reheating temperature when the reheating temperature is high, i.e.,
$\Gamma_I > m_s$ ($\Gamma_I$ : the decay rate of the inflaton ). The
saxion-to-entropy ratio is fixed at the beginning of the saxion
oscillation ($H\sim m_s$), and given by
\begin{eqnarray}
	 \left ( \frac{\rho_s}{s} \right )^{\rm (C)}
	&= \frac{1}{8}T_{{\rm osc}}\left ( \frac{s_i}{M_P} \right )^2 \\
	& \simeq 1.5 \times 10^{-5}~ {\rm GeV}
	\left ( \frac{m_s}{1~{\rm GeV}} \right )^{1/2}
	\left ( \frac{F_a}{10^{12}~{\rm GeV}} \right )^2
	\left ( \frac{s_i}{F_a} \right )^2,  \label{rhosTos}
\end{eqnarray}
where $T_{{\rm osc}}$ denotes the temperature at the beginning of the
saxion oscillation.  On the other hand, if $\Gamma_I < m_s$, the ratio
is fixed at the decay of inflaton $H\sim \Gamma_I$, and the
saxion-entropy ratio is estimated as 
\begin{eqnarray}
	\left ( \frac{\rho_s}{s} \right )^{\rm (C)}
	&= \frac{1}{8}T_R \left ( \frac{s_i}{M_P} \right )^2 \\
	& \simeq 2.1 \times 10^{-9} ~{\rm GeV}
	\left ( \frac{T_R}{10^5~{\rm GeV}} \right )
	\left ( \frac{F_a}{10^{12}~{\rm GeV}} \right )^2
	\left ( \frac{s_i}{F_a} \right )^2.  \label{rhosTR}
\end{eqnarray}

The saxion is also produced by scatterings of
particles in high-temperature plasma.  For $T_R \gtrsim T_D \sim
10^9$GeV $(F_a/10^{11}{\rm GeV})^2$, the saxions are thermalized through
these scattering processes and the abundance is determined
as~\cite{Rajagopal:1990yx}
\begin{equation}
	\left ( \frac{\rho_s}{s} \right )^{\rm (TP)}
	\sim 1.0\times 10^{-3}~{\rm GeV}
	\left ( \frac{m_s}{1~{\rm GeV}} \right ).  \label{rhosTP1}
\end{equation}
For $T_R \lesssim T_D$, this ratio is suppressed by the factor $T_R/T_D$.
The result is
\begin{equation}
	\left ( \frac{\rho_s}{s} \right )^{\rm (TP)}
	\sim 1.0\times 10^{-9}~{\rm GeV}
	\left ( \frac{m_s}{1~{\rm GeV}} \right )
	\left ( \frac{T_R}{10^5~{\rm GeV}} \right )
	\left ( \frac{10^{12}~{\rm GeV}}{F_a} \right )^2. \label{rhosTP2}
\end{equation}
Here we have assumed that thermally produced saxions become
nonrelativistic before they decay.  This assumption is valid for the
parameter regions we are interested in.  We can see that the
contribution from coherent oscillation is proportional to $F_a^2$ while
that from thermal production is proportional to $F_a^{-2}$.  Thus for
small $F_a$ thermal production may be dominant.  Note that this
expression is valid for $T_R\gtrsim m_s$.  Otherwise the saxion cannot
be produced thermally.  As a result, the total saxion abundance is sum of these
two contributions,
\begin{equation}
	\frac{\rho_s}{s} =  
	\left ( \frac{\rho_s}{s} \right )^{\rm (C)}+
	\left ( \frac{\rho_s}{s} \right )^{\rm (TP)}.
\end{equation}

\begin{figure}[tbp]
\begin{center}
    \includegraphics[width=1.0\linewidth]{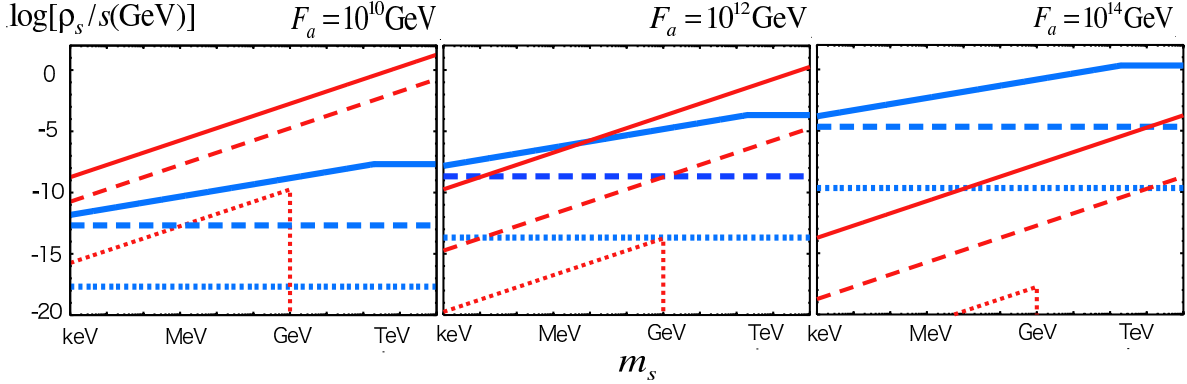}
       \caption{Theoretical predictions for the saxion-to-entropy ratio.
       Thick blue lines represent contribution from the coherent oscillation
       $(\rho_s/s)^{\rm (C)}$ with $s_i \sim F_a$ and thin red ones represent
       thermal contribution $(\rho_s/s)^{\rm (TP)}$.  Solid, dashed and
       dotted lines correspond to $T_R=10^{10}$~GeV,$10^5$~GeV and
       $1$~GeV, respectively.  } \label{fig:energy}
\end{center}
\end{figure}

In Fig.~\ref{fig:energy} we show theoretical predictions for the
saxion-to-entropy ratio with $F_a =10^{10}$~GeV, $10^{12}$~GeV and
$10^{14}$~GeV.\footnote{
Axion overclosure bound ensures $\theta_i^{1.7}
F_a \lesssim 10^{12}$~GeV.  Thus by tuning $\theta_i$, $F_a \sim
10^{14}$~GeV is allowed.  Late-time entropy production also makes such a
large value of $F_a$ viable, although the low reheating temperature $T_R
\lesssim 1$~GeV is needed \cite{Steinhardt:1983ia,Lazarides:1987zf,Kawasaki:1995vt}. }
In the figures the thick blue lines represent contribution from
the coherent oscillation $(\rho_s/s)^{\rm (C)}$ with $s_i \sim F_a$
and thin red ones represent thermal contribution $(\rho_s/s)^{\rm (TP)}$.
Solid, dashed and dotted lines correspond to $T_R=10^{10}$~GeV,$10^5$~GeV and
$1$~GeV, respectively.  We can see that for $F_a \lesssim 10^{12}$~GeV
($F_a \lesssim 10^{10}$~GeV), the contribution
from thermal production dominates at $T_R \gtrsim 10^{10}$~GeV ($T_R
\gtrsim 10^{5}$~GeV).


\subsection{Decay of the saxion}

Since the interaction of the saxion to other particles is suppressed
by the PQ scale $F_a$, it has a long lifetime and decays at
cosmological time scales, which leads to several cosmological effects.
First, let us consider the saxion decay into two axions, $s \to 2a$.  If
we parametrize PQ scalar fields $\Phi_i$ as
\begin{equation}
   \Phi_i = v_i \exp \left [ \frac{q_i \sigma}{\sqrt{2}F_a} \right ],
\end{equation}
where $q_i$ is the PQ charge of the $i$-th PQ field, 
and $F_a = \sqrt{ \sum_i q_i^2 |v_i|^2 }$,
the saxion and axion are identified as $s={\rm Re}[\sigma]$ and  $a={\rm Im}[\sigma]$.
The kinetic term is expanded as
\begin{equation}
   \sum_i |\partial _\mu \Phi_i|^2 \sim 
    \left (1+ \frac{\sqrt 2 f}{F_a}s \right )
    \left ( \frac{1}{2}\partial _\mu a \partial ^\mu a 
     +\frac{1}{2}\partial _\mu s \partial ^\mu s  \right )+ \dots,
\end{equation}
where $f = \sum_i q_i^3 v_i^2/F_a^2$.  From this coupling,
we can estimate the decay rate of the saxion into axions as
\begin{equation}
	\Gamma(s\to 2a) \simeq \frac{f^2}{64\pi}\frac{m_s^3}{F_a^2}.
\end{equation}
If $f \sim 1$ as in many cases including the case with only one PQ
scalar, this is the dominant decay mode of the
saxion~\cite{Chun:1995hc}.
Then the lifetime is given by
\begin{equation}
   \tau_s \simeq 1.3 \times 10^2 f^{-2}~{\rm sec}
    \left ( \frac{1~{\rm GeV}}{m_s} \right )^3
    \left ( \frac{F_a}{10^{12}~{\rm GeV}} \right )^2
\end{equation}
But for the model with superpotential Eq.~(\ref{Wsaxion}), $f$ can be
zero at tree level due to the cancellation if $c_1=c_2$ in
Eq.~(\ref{Vsaxion}).  It is crucial for cosmological arguments
whether the dominant decay mode is into axions or not, because axions
produced in the decay do not interact with other particles and
cosmological constraints can be relaxed if this is the dominant decay
mode.  In this paper, we consider both possibilities $f\sim1$ and $f\sim
0$ and derive cosmological constraints.

Next we consider other modes in which the saxion decays into
standard model particles.  Here we assume the saxion is lighter than
SUSY particles and its decay into SUSY particles is kinematically
forbidden.  Implications of the saxion decay into SUSY particles are
discussed in Sec.~\ref{sec:others}.  For hadronic axion model, the
leading contribution for $m_s \gtrsim 1~{\rm GeV}$ comes from the decay
into two gluons.  The decay rate is estimated as
\begin{equation}
   \Gamma(s \to 2g) \simeq \frac{\alpha_s^2}{64\pi^3}\frac{m_s^3}{F_a^2},
\end{equation}
where $\alpha_s$ denotes the SU(3)$_c$ gauge coupling constant.  The
emitted gluons produce hadron jets which may affect the big bang
necleosynthesis (BBN) as seen in Sec.~\ref{sec:constraints} 

On the other hand, the decay into two photons is always possible, which
has the decay rate,
\begin{equation}
   \Gamma(s \to 2\gamma) \simeq 
    \frac{\kappa^2 \alpha_{\rm EM}^2}{512\pi^3}\frac{m_s^3}{F_a^2},
\end{equation}
where $\alpha_{\rm EM}$ denotes the U(1)$_{\rm EM}$ gauge coupling
constant, and $\kappa $ is a model dependent constant of $O(1)$. 
These photon produced in the decay also bring about
cosmological difficulty.

In the DFSZ axion model, the PQ scalar has tree level coupling with the
ordinary quarks and leptons.  For $m_s > 2m_{ui}(2m_{di})$ where
$u_i$($d_{i}$) denotes the up-type (down-type) quark in the $i$-th
generation ($i=1,2,3$) the saxion decays into a fermion pair with the
decay rate,
\begin{eqnarray}
   \Gamma(s\to u_i \bar u_i) & = & \frac{3}{8\pi} 
    \left ( \frac{2x^{-1}}{x+x^{-1}} \right )^2
	m_s \left ( \frac{m_{ui}}{F_a} \right )^2 
	\left (1- \frac{4m_{ui}^2}{m_s^2} \right )^{3/2},\\
   \Gamma(s\to d_i \bar d_i) & = & \frac{3}{8\pi} 
   \left ( \frac{2x}{x+x^{-1}} \right )^2
	m_s \left ( \frac{m_{di}}{F_a} \right )^2
	\left (1- \frac{4m_{di}^2}{m_s^2} \right )^{3/2}.
\end{eqnarray}
where $x=\tan \beta = \langle H_u \rangle/  \langle H_d \rangle$
(through this paper, we set $x=5$).
Here it should be noticed that for $m_s \lesssim 1$~GeV, the effective
coupling of the saxion with hadrons (mesons) should be used.
However, there are no parameter region
where the saxion decay into mesons has main effects on cosmology in the
following discussion.
Moreover, the decay into the muon pair (see below) gives the same order
of the decay rate for $m_s \gtrsim 210$ MeV
(Decay into mesons are kinematically forbidden for $m_s \lesssim 270$
MeV).
Therefore, for simplicity we neglect the effects of saxion decay into
mesons for $m_s <1$~GeV.  

The saxion coupling with leptons is not
suppressed for the DFSZ axion model, which gives the decay rate
\begin{equation}
   \Gamma(s\to l_i \bar l_i)= \frac{1}{8\pi} 
    \left ( \frac{2x}{x+x^{-1}} \right )^2
	m_s \left ( \frac{m_{li}}{F_a} \right )^2
	\left (1- \frac{4m_{li}^2}{m_s^2} \right )^{3/2}.
\end{equation}
Thus we can see that the saxion decay into heavier fermions is enhanced, 
as long as it is kinematically allowed.
In fact, the decay into fermions may be the dominant mode 
for some mass region even if $f=1$. 

In the KSVZ model, the decay of the saxion into quarks and leptons is
suppressed because the saxion does not directly couple with them. 

Hereafter, we consider the following four typical cases labeled as model (a)-(d).
Model (a) denotes the KSVZ model with $f=1$ and 
model (b) denotes the KSVZ model with $f=0$.
Model (c) denotes the DFSZ model with $f=1$ and 
model (d) denotes the DFSZ model with $f=0$.

\section{Cosmological constraints from saxion}   
\label{sec:constraints}

Given the decay modes of the saxion, we can derive generic
constraints on the saxion density depending on its lifetime and mass.
As we will see, for almost all the mass range ($1$~keV $\lesssim m_s
\lesssim 1$~TeV) the saxion density is bounded from above, although the
upper bounds depend on the cosmological scenario such as the reheating
temperature and the initial displacement of the saxion field.


\subsection{Effective number of neutrinos}

Relativistic particles produced by decaying particles would contribute
to the additional radiation energy density, parametrized by the
increase of the effective number of neutrinos, $\Delta N_\nu$.  The
definition of $N_\nu$ is given through the relation,
\begin{equation}
	\rho_{\rm rad}(T)= \left [ 
	1+ \frac{7}{8}N_\nu \left ( \frac{T_\nu}{T_\gamma} \right )^4
	\right ] \rho_{\gamma}(T_\gamma) ,
\end{equation}
where $\rho_{\rm rad}$ denotes the total relativistic energy density,
$T_{\gamma}$ and $T_\nu$ denote the temperature of the photon and
neutrino.  In the standard model with three species of light neutrinos,
$N_\nu \simeq 3.046$.  But if there exists another species of
relativistic particle, either thermally or nonthermally, it contributes
to the total radiation energy density parametrized by $N_\nu$.  Then the
additional contribution $\Delta N_\nu$ is given as $\Delta N_\nu =
3(\rho_{\rm rad}-\rho_\gamma-\rho_\nu)/\rho_\nu$.  The increase of 
$N_{\nu}$ speeds the Hubble expansion up and causes earlier
freeze-out of the weak interaction, which results in {$^4$He}
overproduction.  The recent analyses of primordial {$^4$He} abundance
\cite{Fukugita:2006xy,Peimbert:2007vm,Izotov:2007ed} are almost
consistent with $N_\nu \sim 3$.
Thus we conservatively adopt $\Delta N_\nu \le 1$ as the BBN
constraint.  Note that this constraint applies to the saxion whose
lifetime is shorter than 1~sec and whose main decay mode is $s\to 2a$
\cite{Choi:1996vz}.

The increase of $\Delta N_\nu$ changes the epoch of the matter-radiation
equality and affects the structure formation of the universe. Thus,
$\Delta N_\nu$ is also constrained from cosmic microwave background
(CMB), galaxy clustering, and Lyman-$\alpha$ forest.  According to the
recent analyses~\cite{Seljak:2006bg,Cirelli:2006kt,Ichikawa:2006vm,
Mangano:2006ur,Hamann:2007pi}, $\Delta N_\nu \gg 1$ is not
favored.\footnote{
It is pointed out that including Lyman-$\alpha$ forest data raises the
best-fit value of $N_\nu$ \cite{Hamann:2007pi}, but $\Delta N_\nu \gg 1$
is still disfavored.  If $\Delta N_\nu$ is close to 1 and $\tau_s > 1$
sec, this may solve the observational discrepancy of $N_\nu$ at BBN and
structure formation \cite{Ichikawa:2007jv}.}  
This constraint applies to the saxion with lifetime $\tau_s \lesssim 10^{13}$~sec.

If the saxion decays mainly into axions, $\Delta N_\nu$ is determined
from the relation,
\begin{equation}
	\frac{\rho_s}{s} \sim 0.34 g_{*s} (T_s)^{-1} \Delta N_\nu T_s,
\end{equation}
where $T_s$ is the temperature at the decay of the saxion and $g_{*s}$
counts the relativistic degrees of freedom.  Thus the requirement
$\Delta N_\nu \le 1$ constrains the saxion abundance as
\begin{equation}
   \frac{\rho_s}{s} \lesssim 3.4\times 10^{-5}~{\rm GeV} 
    \left ( \frac{10}{g_{*s}(T_s)} \right )
    \left ( \frac{T_s}{1~{\rm MeV}} \right ).
    \label{Neffconstraint}
\end{equation}

Note that almost the same constraint is applied even if the saxion decay
into axions is suppressed, if its lifetime is longer than $\sim 1$~sec.  
The reason is as follows.  Roughly speaking,
the constraint $\Delta N_\nu \lesssim 1$ means that the saxion should
not dominate the universe before its decay.  If it dominates the
universe before its decay, a substantial amount of entropy is released
by its decay.  But entropy production after BBN is severely
constrained because the baryon-to-entropy ratio should be unchanged
between BBN and the recombination epoch, as CMB anisotropy measurements and
observed abundances of the light elements
indicates~\cite{Spergel:2006hy}.  
Thus in this case the constraint (\ref{Neffconstraint}) is applied.
On the other hand, entropy production before BBN is possible.\footnote{
See Refs.~\cite{Choi:1996vz,Chun:1999xc} 
for the case of thermal inflation driven by the saxion field trapped at the origin.}
No constraint is imposed in this case if the branching ratio into axions is suppressed
(see the case (b) and (d) of Figs.~\ref{fig:diffuse_10} and \ref{fig:diffuse_12}).
The following analyses and the resulting
constraints on the saxion abundance do not depend on whether the saxion dominates or not.

\subsection{Big-Bang nucleosynthesis}

The saxion with its lifetime $\gtrsim 10^{-2}$~sec may affect
BBN~\cite{Kawasaki:1994af}.  The saxion decays into ordinary particles
either radiatively or hadronically.  If the hadronic decay occurs at
early epoch ($\tau_s \lesssim 10^2~$sec), the main effect on BBN is
$p \leftrightarrow n$ conversion caused by injected pions, which results
in helium overproduction.  At later epoch, photo- and hadro-dissociation
processes of light elements take place efficiently.
When $s \to 2a$ is the dominant decay mode,
the branching ratios into radiation or hadrons are small.
Nevertheless, even a small fraction of the energy
density of the saxion which goes into radiation or hadrons may have
impacts on BBN.  In particular, if hadronic decay modes are open, the
constraint is very stringent.

The constraints from photo-(hadro-)dissociation are approximately
written as
\begin{equation}
   B_r \left (\frac{\rho_s}{s} \right ) \; \lesssim \left\{
	\begin{array}{ll}
	    10^{-6}\, \textrm{--}\, 10^{-14}{\rm\, GeV}&
	     {\rm~~for~~} 10^4{\rm\, sec} \lesssim \tau_s 
	     \lesssim 10^7{\rm\, sec}\\
	    10^{-14} {\rm\, GeV} &
	     {\rm~~for~~} 10^{7}{\rm\, sec} \lesssim \tau_s 
	     \lesssim 10^{12}{\rm\, sec} 
	\end{array}  	
   \right. ,
\end{equation}
for radiative decay, and
\begin{equation}
   B_h \left (\frac{\rho_s}{s} \right ) \;\lesssim \left\{
	\begin{array}{cc}
	    10^{-9}\,\, \textrm{--}\,\, 10^{-13} {\rm\, GeV}&
	     {\rm for~~}1{\rm\, sec} \lesssim \tau_s 
	     \lesssim 10^{4}{\rm\, sec}\\
	    10^{-13}\, \textrm{--}\, 10^{-14} {\rm\, GeV}&
	     {\rm~~~~for~~}10^{4}{\rm\, sec} \lesssim \tau_s 
	     \lesssim 10^{12}{\rm\, sec} 
	\end{array}	  	
   \right. ,
\end{equation}
for hadronic decay, where $B_r$ and $B_h$ denote the radiative and
hadronic branching ratios, respectively (here $B_r$ includes the
hadronic decay modes).  Note that if the injected photon energy (which
is equal to the half of the saxion mass) is smaller than the threshold
energy to destroy the light elements especially {$^4$He}, which is
typically $O(10)$ MeV, the photo- and hadro-dissociation constraints are
much weakened.  In particular BBN constraints are neglected for $m_s
\lesssim 4.5$ MeV, which corresponds to the threshold energy for the
process ${\rm D} + \gamma \to n+p$.

\subsection{Cosmic microwave background}

The saxion with lifetime $10^{6}$~sec $\lesssim \tau_s \lesssim
10^{13}$~sec may affect the blackbody spectrum of CMB.  Since preserving
the blackbody spectrum requires the photon number-violating processes 
such as double-Compton scattering, 
to maintain thermal equilibrium between photons and electrons,
photons injected in the decay distort the
CMB spectrum at $t\gtrsim 10^6~$sec when the double-Compton scattering
becomes inefficient.  The distortion is characterized by the chemical
potential $\mu$ at $t\lesssim 10^9$~sec when the energy transfer by the
Compton scattering is efficient, and Compton $y$-parameter at later
epoch, which characterizes the deviation of the CMB spectrum from
thermal distribution due to the inverse Compton scattering by high
energy electrons.  They are constrained from COBE FIRAS measurement as
$|\mu| \lesssim 9\times 10^{-5}$ and $y \lesssim 1.2\times
10^{-5}$~\cite{Fixsen:1996nj}.  $\mu$ and $y$ are related to the
injected photon energy $\delta\rho_{\gamma}$
as~\cite{Ellis:1990nb,Hu:1993gc}
\begin{equation}
	\frac{\delta \rho_\gamma}{\rho_\gamma}\sim 0.714\mu,
\end{equation}
for $10^6$~sec $\lesssim \tau_s \lesssim 10^9$~sec, and
\begin{equation}
	\frac{\delta \rho_\gamma}{\rho_\gamma}\sim 4y,
\end{equation}
for $10^9$~sec $\lesssim \tau_s \lesssim 10^{13}$~sec.
This in turn constrains the saxion energy density depending on its 
branching ratio into radiation $B_r$, as
\begin{equation}
   B_r\left (\frac{\rho_s}{s} \right ) \lesssim 
    \left \{ 
     \begin{array}{ll}
        9.0\times 10^{-13}~{\rm GeV}
	 \left ( \frac{10^{9}~{\rm sec}}{\tau_s} \right )^{1/2} 
	 ~&(10^6~{\rm sec} \lesssim \tau_s \lesssim 10^9~{\rm sec})\\
        6.7\times 10^{-13}~{\rm GeV}
	 \left ( \frac{10^{9}~{\rm sec}}{\tau_s} \right )^{1/2} 
	 ~&(10^9~{\rm sec} \lesssim \tau_s \lesssim 10^{13}~{\rm sec})
     \end{array}
   \right. .
\end{equation}

\subsection{Diffuse X($\gamma$)-ray background}

The two photon decay of the saxion with lifetime longer than $\sim
10^{13}$~sec may contribute to diffuse X$(\gamma)$-ray background.  The
mass of the saxion which has such a long lifetime is typically smaller
than 1~GeV.
The photon with energy 1~keV$\lesssim E_\gamma \lesssim 1$~TeV is
transparent against the scattering with cosmic background photons and
intergalactic medium, and hence such decay-produced photons freely
propagate through the universe and can be observed as diffuse background
photons \cite{Chen:2003gz}.

The flux of the photons from the decay of the saxion is calculated
as~\cite{Kawasaki:1997ah}
\begin{equation}
   F_\gamma(E) = \frac{E}{4\pi} \int_0^{t_0}dt 
    \frac{B_\gamma n_s(z)}{\tau_s}(1+z)^{-3}
    \frac{dE^\prime}{dE}2\delta \left ( E^\prime - \frac{m_s}{2} \right ),
\end{equation}
where $B_\gamma$ denotes the branching ratio into two photons and
$n_s(z)$ is the number density of the saxion at the redshift $z$.
$E^\prime $ is the energy of the photon at the instant of production and
$E$ is the present redshifted energy, the relation between them is given
by $E^\prime = (1+z)E$.  Under the assumption of the flat universe
($\Omega_\Lambda+\Omega_m =1$), this expression can be integrated
yielding
\begin{eqnarray}
   F_\gamma(E)=&\frac{B_\gamma n_{s 0}}{2\pi \tau_s H_0} 
    g\left ( \frac{m_s}{2E} \right ) \nonumber \\
   &\times \exp \left [ \frac{1}{3H_0 \tau_s \sqrt{\Omega_\Lambda}}
		 \ln \frac{ \left ( \sqrt {\Omega_\Lambda} 
			     g\left (\frac{m_s}{2E}\right ) -1 \right )
			     \left ( \sqrt {\Omega_\Lambda}  +1 \right ) }
			     { \left ( \sqrt {\Omega_\Lambda} 
				g\left (\frac{m_s}{2E}\right ) +1 \right )
				\left ( \sqrt {\Omega_\Lambda} -1 \right )} 
		\right ]
	\label{Xspectrum}
\end{eqnarray}
where $n_{s0}$ denotes the present number density of the saxion,
$H_0$ denotes the present Hubble constant, 
and
\begin{equation}
	g(x) = \left [ \Omega_\Lambda + \Omega_m x^3 \right ]^{-1/2}.
\end{equation}

On the other hand, the observed photon flux in the range 1~keV $\lesssim
E \lesssim 100$~GeV is roughly given as
\begin{equation}
 F_{\gamma {\rm obs}}(E) \sim 
  \left \{ 
   \begin{array}{ll}
    8\left ( \frac{E}{{\rm keV}} \right )^{-0.4} ~~~~~
     &(0.2~{\rm keV} \lesssim E \lesssim 25~{\rm keV})\\
    57\times 10^{-4} \left ( \frac{E}{{\rm MeV}} \right )^{-1.6} 
     &(25~{\rm keV} \lesssim E \lesssim 4~{\rm MeV})\\
    17\times10^{-6}\left ( \frac{E}{100~{\rm MeV}} \right )^{-1.1} 
     &(4~{\rm MeV} \lesssim E \lesssim 120~{\rm GeV}),
   \end{array}
  \right.
\end{equation}
in the unit of cm$^{-2}$ sec$^{-1}$ sr$^{-1}$, from the observations of
ASCA \cite{Gendreau:1995}, HEAO1 \cite{Kinzer:1997}, COMPTEL
\cite{Kappadath:1996}, and EGRET \cite{Sreekumar:1997un}.  Thus, from the
requirement $F_\gamma < F_{\gamma {\rm obs}}$, the tight constraint on
the saxion density is derived.  Note that even for the saxion lifetime
longer than the age of the universe, the small fraction of the
decayed saxion at $t<t_0$ contributes to the diffuse background and its
abundance is limited.

To estimate this constraint, let us consider the case with $\tau_s > t_0$.
In this case, the X$(\gamma)$-ray spectrum of photons from the saxion decays
has the maximum at $E_{\rm max}=m_s/2$, where the flux is simplified as
$F_\gamma(E_{\rm max})=B_\gamma n_{s0}/(2\pi \tau_s H_0)$.  Then from
the condition $F_\gamma(E_{\rm max}) < F_{\gamma {\rm obs}}(E_{\rm
max})$, we obtain a constraint,
\begin{eqnarray}
 B_\gamma \left (\frac{\rho_s}{s} \right )
  & \lesssim 2\pi \frac{m_s \tau_s H_0}{s_0} 
  F_{\gamma {\rm obs}}\left ( \frac{m_s}{2} \right )  \nonumber \\
 & \sim 2.4h\times 10^{-18}~{\rm GeV}
  \left ( \frac{m_s}{1~{\rm MeV}} \right ) \nonumber \\
 &~~~~\times \left ( \frac{\tau_s}{10^{18}~{\rm sec}} \right )
  \left ( \frac{F_{\gamma {\rm obs}}(m_s/2)}
   {10^{-2}~{\rm cm^{-2}sec^{-1} }} \right ),
\end{eqnarray}
where $s_0$ denotes the present entropy density.  For the case with
$\tau_s < t_0$, the photon energy which gives the flux maximum deviates
from $E=m_s/2$ due to the Hubble expansion.  For $\tau_s \ll t_0$, it is
given by $E_{\rm max}=(m_s/2)(3H_0\tau_s \sqrt{\Omega_m}/2)^{2/3}$.
This leads to the constraint
\begin{eqnarray}
 B_\gamma \left (\frac{\rho_s}{s} \right )
  & \lesssim \frac{4\pi}{3} \frac{m_s}{s_0} 
  F_{\gamma {\rm obs}}\left ( E_{\rm max} \right )  \nonumber \\
 & \sim 4.8 \times 10^{-19}~{\rm GeV}
  \left ( \frac{m_s}{1~{\rm MeV}} \right ) 
  \left ( \frac{F_{\gamma {\rm obs}}(E_{\rm max})}
   {10^{-2}~{\rm cm^{-2}sec^{-1} }} \right ).
\end{eqnarray}

\subsection{Reionization}

If the saxion decays after recombination era and the injected photon energy
is relatively small ($m_s \lesssim O(1)$~keV-$O(1)$~MeV and $10^{13}$~sec
$\lesssim \tau_s$), redshifted photons may leave the transparency
window until the present epoch \cite{Chen:2003gz}.  Then, emitted
photons interact with and ionize the intergalactic medium (IGM), and
they contribute as an additional source of the reionization.  If this
contribution is too large, the optical depth to the last scattering
surface is too large to be consistent with the WMAP
data~\cite{Spergel:2006hy}.  Here we apply the results from
Refs.~\cite{Chen:2003gz,Zhang:2007sc}, simply assuming that if the
decay-produced photon leaves the transparency window, one-third of the
photon energy is converted to the ionization of the IGM (the remaining
goes to the excitation and heating of the IGM).  According to
Refs.~\cite{Chen:2003gz,Zhang:2007sc}, this is a good approximation when
the decay occurs before the reionization due to astrophysical objects
takes place and most of hydrogen atoms exist in the form of neutral
state.
(The Gunn-Peterson test indicates that the reionization
occurred at $z\sim6$ \cite{Fan:2005es}.)
The constraint on the saxion density can be written as
\begin{equation}
 B_r  \left (\frac{\rho_s}{s} \right ) \lesssim  
  \left (\frac{\rho_s}{s} \right )_{\rm bound},
\end{equation}
where $ \left (\rho_s/s \right )_{\rm bound}$ can be read off from
Fig.~2 of Ref.~\cite{Zhang:2007sc}.  For example, for $\tau_s \gtrsim t_0$,
it is given by
\begin{equation}
 \left (\frac{\rho_s}{s} \right )_{\rm bound} 
  \simeq 4.3 \times 10^{-17}~{\rm GeV}
  \left ( \frac{\tau_s}{10^{18}~{\rm sec}} \right )
  \left ( \frac{\Omega_b h^2}{0.022} \right ),
\end{equation}
where $\Omega_b$ denotes the density parameter of the baryonic matter.
This constraint is complementary to the diffuse X($\gamma$)-ray limit.

\subsection{Present matter density limit}  
\label{sec:matterdensity}

For the saxion with its lifetime $\tau_s > t_0$, its energy density
contributes to the dark matter of the universe, and hence the saxion
density should be less than the observed matter density, $\Omega_{s}h^2
\lesssim \Omega_m h^2$.  In terms of the saxion-to-entropy ratio, this
is written as
\begin{equation}
 \frac{\rho_s}{s} \lesssim 4.7\times 10^{-10}~{\rm GeV}
  \left ( \frac{\Omega_m h^2}{0.13}\right ).
\end{equation}

\subsection{LSP overproduction}  \label{sec:LSP}

If the saxion mass is larger than about 1 TeV, the saxion can decay into
SUSY particles.  Here we suppose that the LSP is the lightest
neutralino.  The decay into SUSY particles were investigated in detail
in Ref.~\cite{Endo:2006ix} and it was found that decay into gauginos has
roughly the same branching ratio as that into gauge bosons.  Thus we
should be careful about LSP overproduction from the saxion decay.  The
resultant abundance of the LSP depends on $T_s$,
and for $T_s \gtrsim m_{\rm LSP}/20$ LSPs produced
from the saxion decay are thermalized and have the same abundance as
that expected in the standard thermal relic scenario of the LSP dark
matter.  In this case no upper bound on the saxion abundance is imposed.
On the other hand, if $T_s \lesssim m_{\rm LSP}/20$,
the abundance of the LSP is given by
\begin{equation}
	\frac{\rho_{\rm LSP}}{s} \simeq 
	\left \{ 
	   \begin{array}{ll}
	B_s\displaystyle \frac{2m_{\rm LSP}}{m_s} \frac{\rho_s}{s}
	+ \frac{\rho_{\rm LSP}^{\rm thermal}}{s} ~~~{\rm for}~~
	n_{\rm LSP}(T_s) \langle \sigma v \rangle < H(T_s),  \label{NTLSP} \\
	\displaystyle \sqrt {\frac{45}{8\pi^2 g_*(T_s)} } 
	     \frac{m_{\rm LSP}}{\langle \sigma v \rangle T_s M_P}~~~{\rm for}~~
	n_{\rm LSP}(T_s) \langle \sigma v \rangle > H(T_s),
	\end{array}
	\right.
\end{equation}
where $B_s$ denotes the branching ratio of the saxion into SUSY
particles, $\langle \sigma v \rangle$ denotes the thermally averaged
annihilation cross section of the LSP,
and $\rho_{\rm LSP}^{\rm thermal}$ denotes the contribution from thermal relic LSPs
taking account of the dilution from the saxion decay.
The LSP number density immediately after the saxion decay $n_{\rm LSP}(T_s)$
is defined as $n_{\rm LSP}=2B_s\rho_s(T_s)/m_s$.
For deriving the constraint,
we ignored the contribution to the LSP production from thermal scattering processes.
Moreover, the second line of Eq.~(\ref{NTLSP}) always results in overproduction of LSPs
with the annihilation cross section for ordinary neutralino dark matter.
Thus for deriving the constraint, we consider only the first term of the 
first line of the right hand side of Eq.~(\ref{NTLSP}).
The bound can be written in the form
\begin{equation}
	\frac{\rho_s}{s} \lesssim 2.4\times 10^{-10}~{\rm GeV}
	\left ( \frac{m_s}{m_{\rm LSP}} \right )
	\left ( \frac{\Omega_m h^2}{0.13}\right ),
\end{equation}
for $m_s\gtrsim 1$~TeV and $T_s \lesssim m_{\rm LSP}/20$. 
This constraint can be relaxed if the annihilation cross section of
the LSP is significantly large.
We will revisit this issue in Sec.~\ref{sec:others}.
Hereafter we set $m_{\rm LSP}=500$~GeV as a reference value.

\begin{figure}[htbp]
 \begin{center}
  \includegraphics[width=1.0\linewidth]{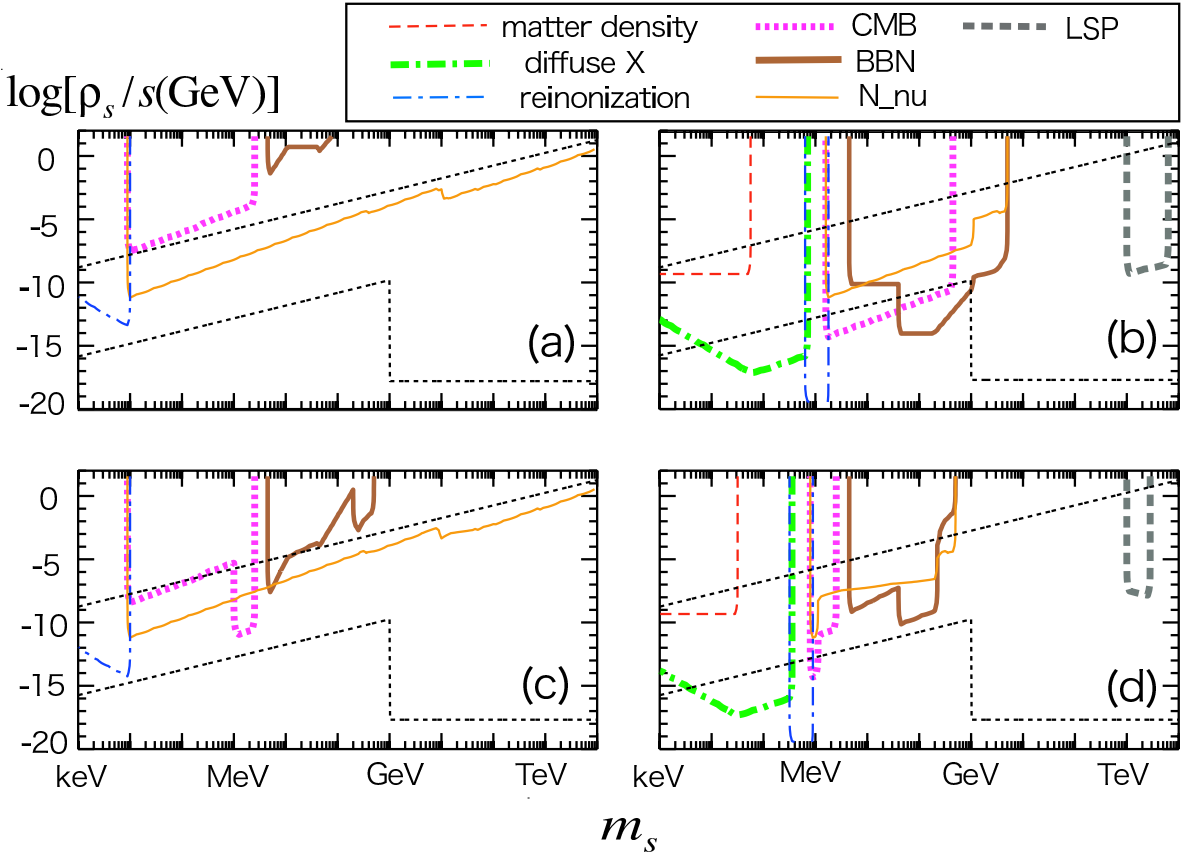} 
  \caption{ Various cosmological constraints on the saxion abundance for
  $F_a = 10^{10}~$GeV.  Thin dotted black lines represent theoretical prediction
  $\rho_s/s = (\rho_s/s)^{\rm (C)}+(\rho_s/s)^{\rm (TP)}$ for $T_R =
  10^{10}$~GeV (upper) and $T_R =1$~GeV (lower) with $s_i = F_a$.  Four
  panels correspond to different models.  Model (a) : KSVZ with $f=1$,
  model (b) : KSVZ with $f=0$, model (c) : DFSZ with $f=1$, model (d) : DFSZ with $f=0$.}
  \label{fig:diffuse_10}
 \end{center}
\end{figure}

\begin{figure}[htbp]
 \begin{center}
  \includegraphics[width=1.0\linewidth]{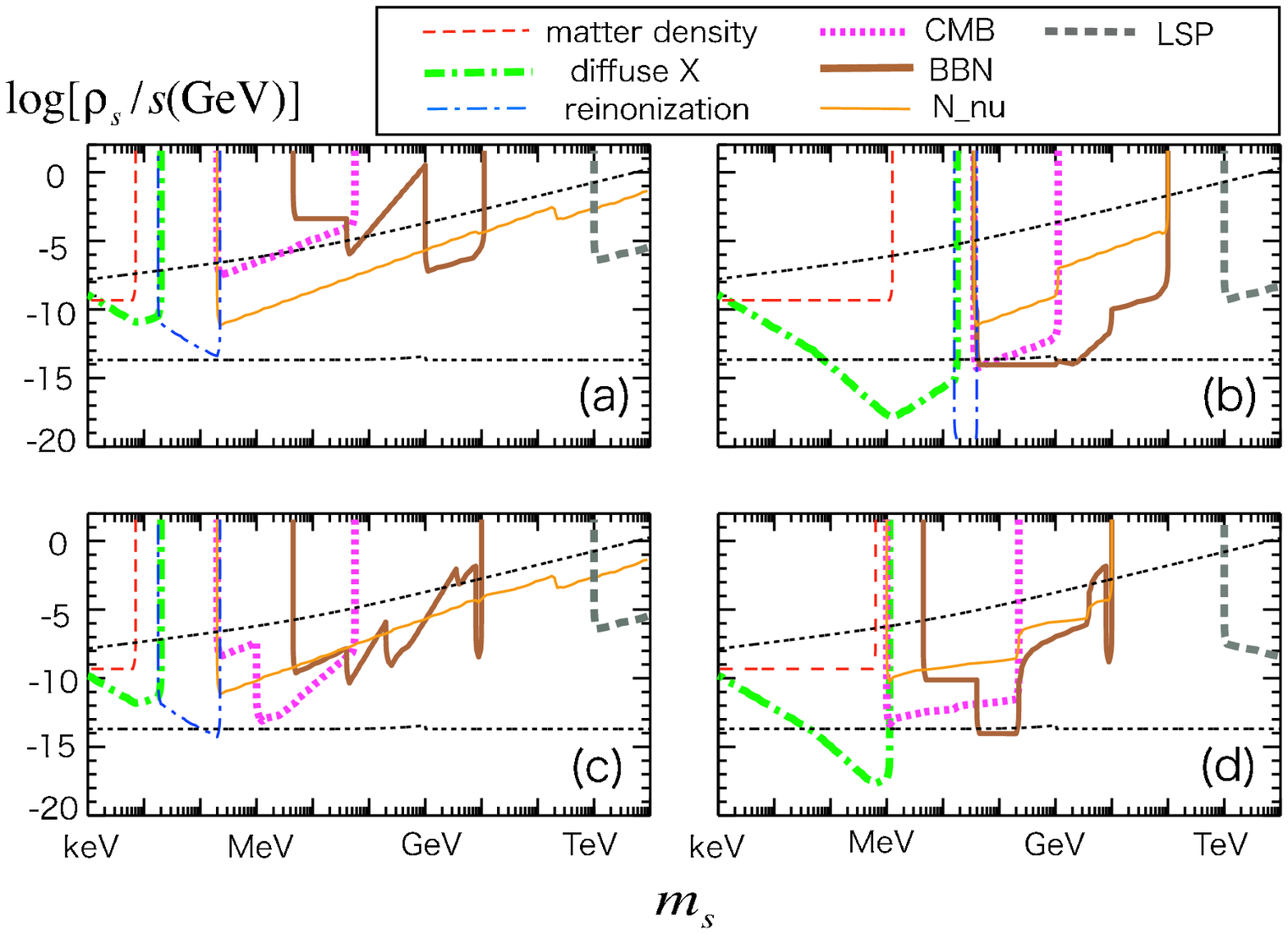}
  \caption{ Same as Fig.~\ref{fig:diffuse_10}, but for
		$F_a=10^{12}$~GeV.  } \label{fig:diffuse_12}
 \end{center}
\end{figure}

\begin{figure}[htbp]
 \begin{center}
  \includegraphics[width=1.0\linewidth]{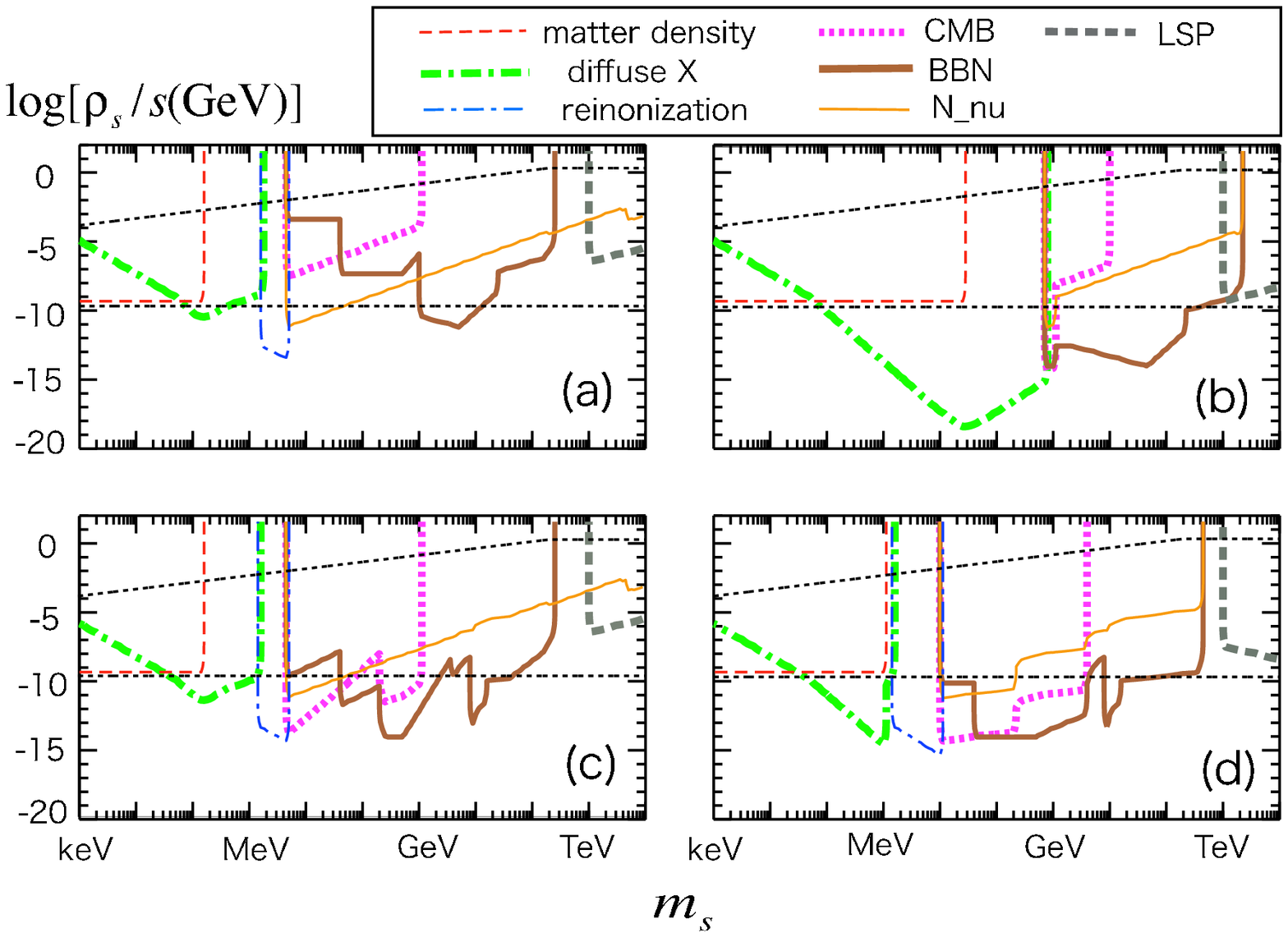}
  \caption{ Same as Fig.~\ref{fig:diffuse_10}, but for $F_a=10^{14}$~GeV.  }
		\label{fig:diffuse_14}
 \end{center}
\end{figure}

Including all of these constraints, we can derive general upper bounds
on the saxion-to-entropy ratio as a function of the saxion mass $m_s$
for models (a)-(d).  In Fig.~\ref{fig:diffuse_10}-\ref{fig:diffuse_14},
we show the results with $F_a=10^{10},10^{12}$ and $10^{14}$~GeV,
respectively. 
In each panel, the orange line represents the bound from $\Delta N_\nu
\lesssim 1$, the thick-solid brown line represents the bound from BBN, the thick-dotted purple
line represents the bound from CMB, the thick-dot-dashed green line represents the bound
from diffuse X($\gamma$)-ray background, the thin-dot-dashed blue line represents the
bound from reionization, the thin-dashed red line represents limit from the present
matter density, and the thick-dashed gray line represents LSP overproduction limit
from the saxion decay.  We also show the theoretical prediction
for the saxion energy density in the
figures for $T_R=10^{10}$~GeV and 1~GeV by thin-dotted black lines.


\section{Constraints from axino}

So far, we have ignored the cosmological effects of axino, the fermionic
superpartner of axion.  The mass of axino is model dependent, but it can
be as heavy as the gravitino mass, $m_{3/2}$
\cite{Goto:1991gq,Chun:1995hc,Chun:1992zk}.  For example, in the model of
Eq.~(\ref{Wsaxion}), the axino mass is estimated as $m_{\tilde a}\sim
|\lambda X|$.  As noted earlier, taking into account the $A$-term
potential like $V_A \sim m_{3/2} \lambda X F_a^2+{\rm h.c.}$,
$X$ can have the VEV of the order of $m_{3/2}$.  Thus in this model the axino
mass is naturally expected to be $m_{3/2}$.  
Hereafter for simplicity we assume $m_{\tilde a} \sim m_{3/2}$.

Axinos and gravitinos are produced through scatterings of particles in
thermal bath.  First, we assume either of them is the LSP, and hence their
thermally produced abundance must not exceed the present abundance of
the dark matter~\cite{Moroi:1993mb}.\footnote{%
As long as $m_s \sim m_{3/2} \sim m_{\tilde a}$, which of them is
lighter is not relevant.  But if $m_{\tilde a} \ll m_{3/2} $ and
$m_{3/2} \gtrsim$ 1~GeV, thermally produced gravitinos can decay into
axinos within the present age of the universe and the constraint is
relaxed~\cite{Asaka:2000ew}.  }
The abundance of thermally produced axinos is calculated
as~\cite{Covi:2001nw}\footnote{
Here we assume there is no entropy production after the reheating ends.
If the saxion dominates the universe and decays before BBN, 
the axino and gravitino abundance can be reduced.
}
\begin{equation}
 \frac{\rho_{\tilde a}}{s}
  \simeq 2.0 \times 10^{-7}g_s^6~{\rm GeV}   
  \left ( \frac{m_{\tilde a}}{1 {\rm\, GeV}} \right )
  \left ( \frac{10^{12}{\rm\,GeV}}{F_a} \right )^{2} 
  \left ( \frac{T_R}{10^{6}{\rm\,GeV}} \right ),
  \label{TPaxino_omega}
\end{equation}
where $g_s$ is the QCD gauge coupling constant.  Thus for large
$m_{\tilde a} (\sim m_{s})$, the constraint on the reheating temperature
becomes stringent.
Note that axino thermal production for $T_R < 1$TeV is negligible,
because SUSY particles are not produced efficiently for such a low  
reheating temperature.
  
Next, if the axino mass is larger than about 1 TeV, the
axino can decay into SUSY particles.  This LSP abundance 
depends on the temperature at the decay of the axino, $T_{\tilde a}$
similar to the case of the saxion decay.  If $T_{\tilde a}\gtrsim m_{\rm
LSP}/20$, LSPs produced from the axino decay are thermalized and the
standard thermal relic scenario of the LSP dark matter is maintained.  If
$T_{\tilde a}\lesssim m_{\rm LSP}/20$, the LSP abundance is determined
by Eq.~(\ref{NTLSP}) after replacing $T_s$ and $\rho_s/s$ with
$T_{\tilde a}$ and $\rho_{\tilde a}/s$.

On the other hand, the thermally produced gravitinos have the abundance
as~\cite{Bolz:2000fu}
\begin{equation}
 \frac{\rho_{3/2}^\mathrm{TP}}{s} 
  \simeq 6.3\times 10^{-11}~{\rm GeV} 
  \left ( \frac{1{\rm\,GeV}}{m_{3/2}} \right )
  \left ( \frac{m_{\tilde g}}{1 {\rm\,TeV}} \right )^{2} 
  \left ( \frac{T_R}{10^6{\rm\,GeV}} \right )  \label{TPgravitino}
\end{equation}
for $m_{3/2} \ll m_{\tilde g}$ where $m_{\tilde g}$ denotes the mass of
the gluino (here the logarithmic dependence on $T_R$ is omitted).
Contrary to the axino, the constraint becomes severer when $m_{3/2}
(\sim m_s)$ becomes smaller.  Note that for $m_{3/2} \lesssim 1$~keV,
gravitinos get thermalized and their abundance becomes independent of
the reheating temperature.  But for 16~eV $\lesssim m_{3/2} \lesssim
1$~keV they contribute to the dark matter density as a hot component due
to their long free-streaming length, and such a contribution is
constrained from cosmological observations in particular Lyman-$\alpha$
forest data~\cite{Viel:2005qj}.  Thus the gravitino mass in this region
is strongly disfavored.  
(The case of ultra-light gravitino $m_{3/2} \lesssim 16$~eV will be 
mentioned later.)
In addition to the constraint from the present
matter density, there may be another constraint coming from the
late-decay of SUSY particles into gravitinos or axinos, which may affect
BBN.  But the constraint is quite model dependent, and hence we do not
consider it here.

\begin{figure}[tbp]
 \begin{center}
  \includegraphics[width=1.0\linewidth]{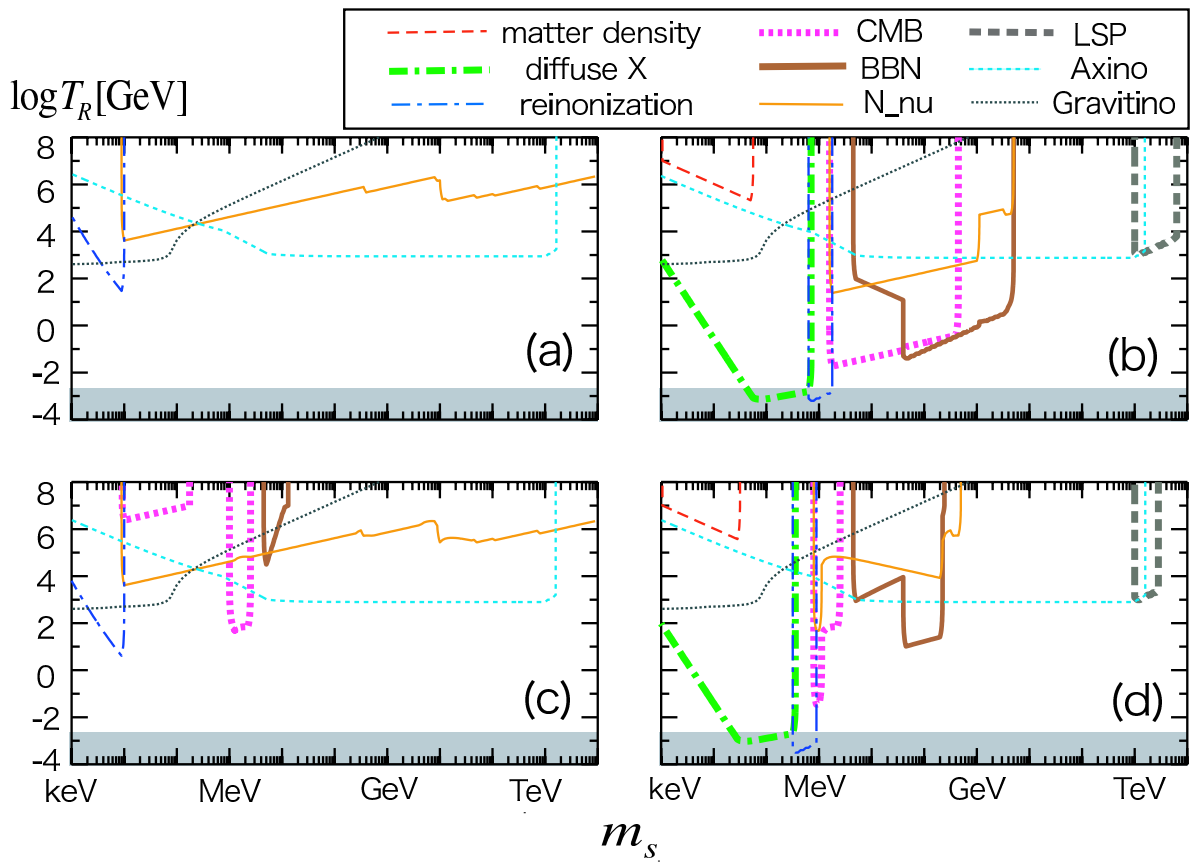} 
  \caption{Upper bounds on the reheating temperature $T_R$ for each
  model with $F_a=10^{10}$~GeV.  The initial amplitude of the saxion is
  assumed to be $s_i \sim F_a$.  The thin-short-dashed light blue line represents the
  bound from axino thermal production, and thin-dotted black line represents the
  bound from gravitino thermal production.  The shaded region
  contradicts with the lowest possible reheating temperature
  \cite{Kawasaki:1999na}.  The other lines are the same as
  Fig.~\ref{fig:diffuse_10}.
  Four panels correspond to different models.  Model (a) : KSVZ with $f=1$,
  model (b) : KSVZ with $f=0$, model (c) : DFSZ with $f=1$, model (d) : DFSZ with $f=0$.}  
  \label{fig:msTR_10}
 \end{center}
\end{figure}

\begin{figure}[htbp]
 \begin{center}
  \includegraphics[width=1.0\linewidth]{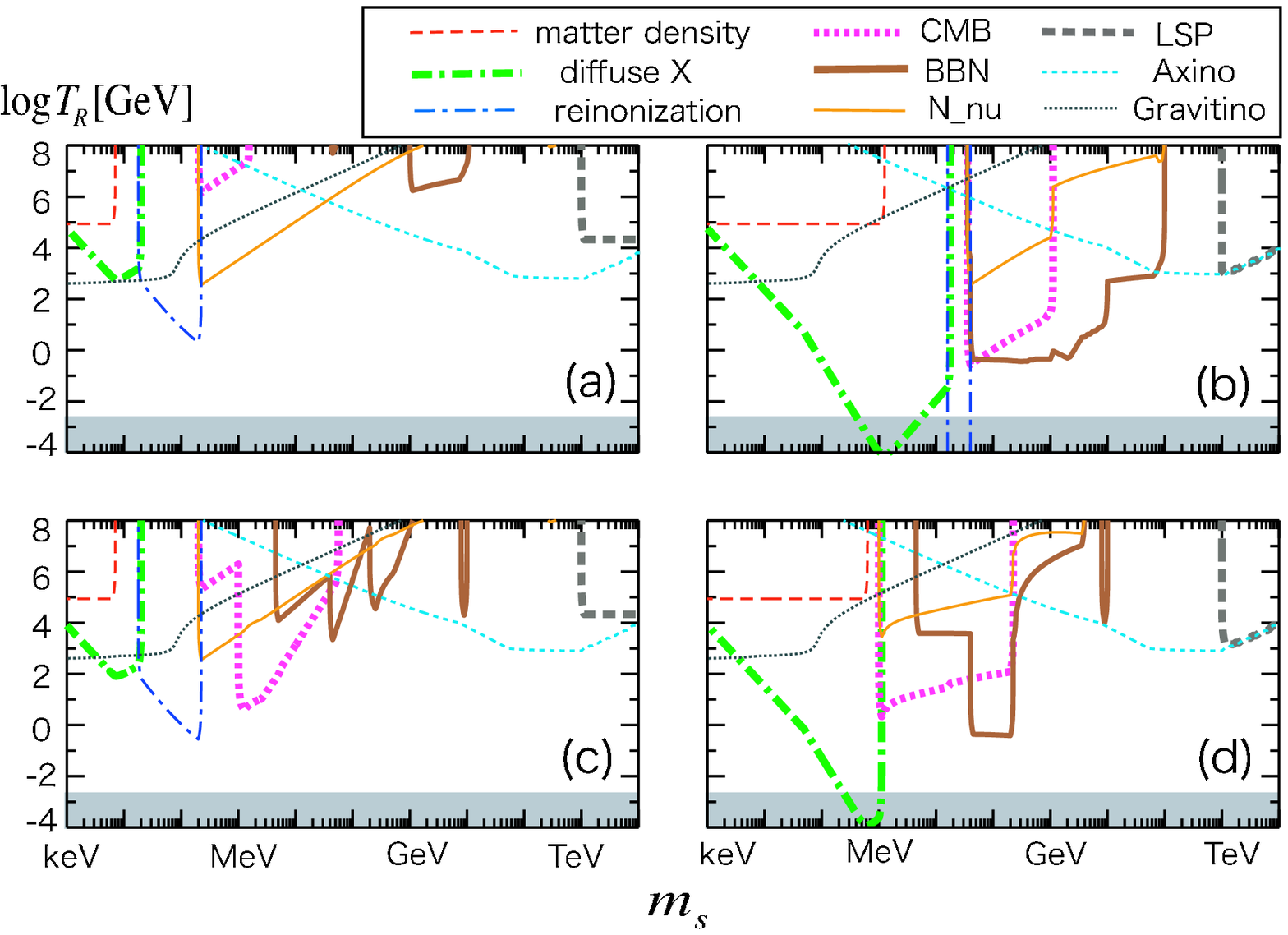} 
  \caption{Same as Fig.~\ref{fig:msTR_10}, but for $F_a=10^{12}$~GeV.  }
		\label{fig:msTR_12}
 \end{center}
\end{figure}

\begin{figure}[htbp]
 \begin{center}
  \includegraphics[width=1.0\linewidth]{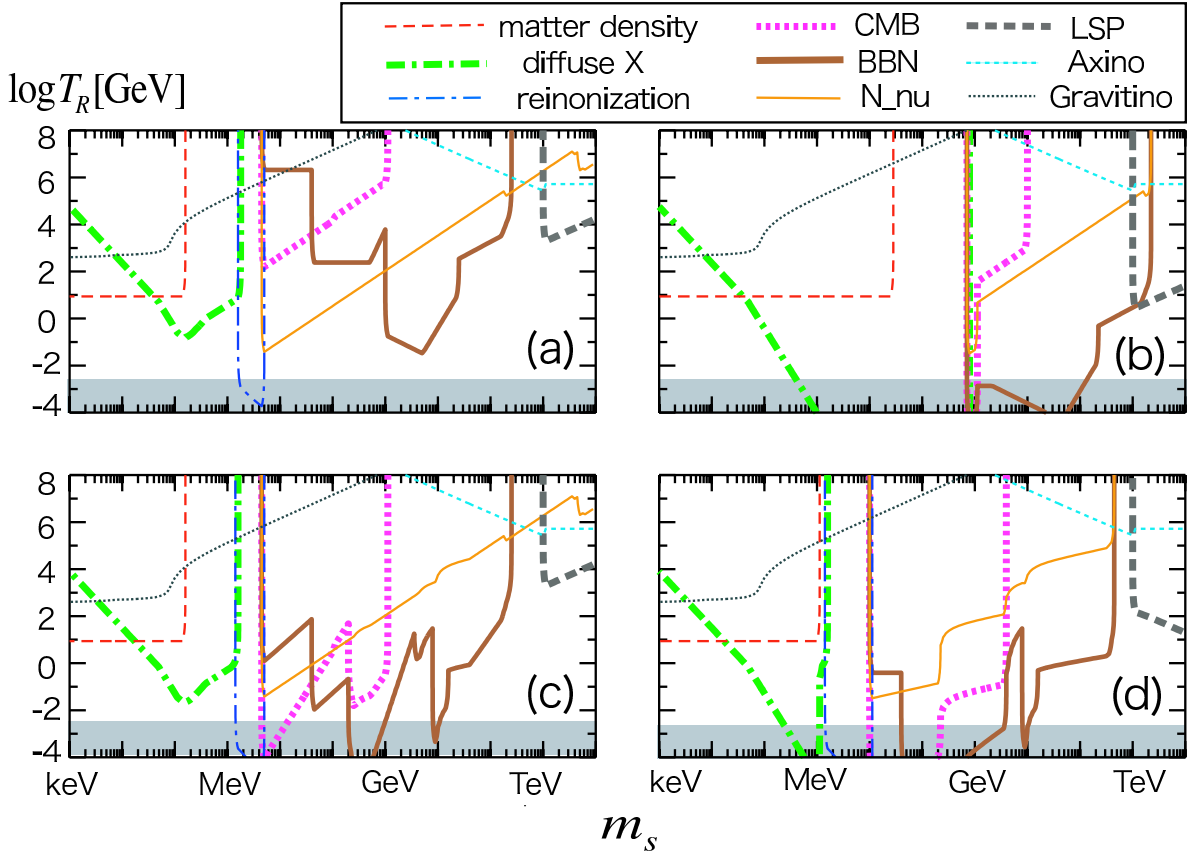}
  \caption{Same as Fig.~\ref{fig:msTR_10}, but for $F_a=10^{14}$~GeV.  }
		\label{fig:msTR_14}
 \end{center}
\end{figure}

Including those constraints from the gravitino and axino, we derive the
upper bounds on the reheating temperature for each saxion mass and show
them in Figs.~\ref{fig:msTR_10}-\ref{fig:msTR_14}.  In these figures, we
have assumed that the initial amplitude of the saxion is given by $s_i
\sim F_a$.  The thin-short-dashed light blue line represents the bound from axino thermal
production, and thin-dotted black line represents the bound from gravitino thermal
production.  The other lines are the same as
Fig.~\ref{fig:diffuse_10}.  Because the saxion-to-entropy ratio
$(\rho_s/s)^{\rm (C)}$ is proportional to $(T_R s_i^2)$, the constraints
on $T_R$ scale as $s_i^{-2}$.
However, it should be noted that
although the axino constraint is stringent for relatively large $m_s$
as can be seen from these figures,
it may be significantly relaxed if the axino is much lighter than the  
gravitino.


\section{Dark matter candidates}   
\label{sec:others}

We have seen that the very stringent bound on the reheating temperature
is imposed for wide range of the saxion mass.  It is typically stronger
than the usual upper bound from the gravitino overproduction.  It has
some implications to the dark matter candidates.  For example, it
invalidates the gravitino dark matter for wide parameter regions.  In
this section we summarize the dark matter candidate in SUSY axion
models.

\subsection{$m_s\lesssim 1$~TeV}

First consider the case where the gravitino (axino) is the LSP.  In our
model, we assume the axino has the mass comparable to the gravitino, and
hence the axino (gravitino) is the NLSP.  Which of them is the lighter is
not important because both have the similar properties.  The saxion also
has the mass comparable to the gravitino.  The stability of the saxion
is not ensured by the $R$-parity, but since its decay rate is suppressed
by the PQ scale $F_a$, the saxion lifetime can exceed the present age of
the universe.  If this is the case, the saxion can be the dark matter.
In addition, if $\theta_i ^{1.7}F_a \sim 10^{12}$~GeV, the axion can
also play a roll of the dark matter.  Therefore, we have four candidates
for the dark matter, i.e. gravitino, axino, axion and saxion.  However,
the saxion dark matter is possible only when $F_a \sim 10^{14}$~GeV and
1~keV$\lesssim m_s \lesssim 10$~keV, as can be seen from
Fig.~\ref{fig:msTR_14}, and hence is less attractive candidate.  For
$10^{10}$~GeV $\lesssim F_a \lesssim 10^{12}$~GeV, the axino dark matter
is allowed for wide parameter regions, but the gravitino dark matter is
excluded except for $m_s \lesssim100$~keV, as can be seen from
Fig.~\ref{fig:msTR_10} and Fig.~\ref{fig:msTR_12}.  On the other hand,
for $F_a \gtrsim 10^{14}$~GeV, the axino or gravitino dark matter is
almost impossible and the axion is the most viable dark matter
candidate, although other cosmological constraints are severe
(Fig.~\ref{fig:msTR_14}).

To summarize, for $m_s\lesssim 1$~TeV, the axino is a good dark
matter candidate for $F_a \lesssim 10^{12}$~GeV, and the axion may be
dark matter candidate for $F_a \gtrsim 10^{12}$~GeV.

\subsection{$m_s \gtrsim 1$~TeV}  

For larger $m_s(\simeq m_{3/2})$, the saxion decay mode into SUSY
particles opens.  Then the LSP is assumed to be the lightest neutralino.  As
discussed in Sec.~\ref{sec:LSP}, the abundance is given by
Eq.~(\ref{NTLSP}) if the decay occurs after the freeze-out of the LSP.
For the neutralino which has small annihilation cross section, very low
reheating temperature is needed to obtain the correct abundance of the
dark matter as can be seen from
Figs.~\ref{fig:msTR_10}-\ref{fig:msTR_14}.  However, for the neutralino
with larger annihilation cross section such as wino- or higgsino-like
LSP, the abundance may be significantly reduced and they can become the
dark matter independent of the reheating temperature~\cite{Endo:2006ix}.
This is easily seen from Eq.~(\ref{NTLSP}),
which describes that the resulting abundance is bounded from above.
Thus for the saxion mass larger than 1~TeV, these non-thermally produced
neutralinos may be dark matter.  On the other hand, if $T_s$ is larger
than the freeze-out temperature of the LSP $T_f \sim m_{\rm LSP}/20$,
the standard thermal relic scenario holds and the saxion has no impact
on cosmology unless it dominates the universe before the decay.  If the
saxion once dominates the universe, it must not decay into axions
mainly, since otherwise $\Delta N_\nu \gg 1$ holds at BBN and
contradicts with the observation.

Therefore, for $m_s\gtrsim 1$~TeV, the lightest neutralino produced
either thermally or non-thermally is likely dark matter depending on
$m_s$ and $F_a$.  Of course, the axion is also a good candidate for the
dark matter for $F_a \gtrsim 10^{12}~$GeV.

\section{Ultra-light gravitino scenario}   
\label{sec:light}

We have seen that for almost all the mass range of the gravitino, 
the reheating temperature is severely constrained.
However, for an ultra-light gravitino with mass of the order of 1-10~eV, 
gravitinos are thermalized and their abundance is sufficiently lower than 
the dark matter abundance,
and hence the reheating temperature is not constrained from gravitino overproduction.
Thus thermal leptogenesis using right-handed neutrino \cite{Fukugita:1986hr}, 
which requires $T_R \gtrsim 10^9$~GeV, may be possible.

\begin{figure}[htbp]
 \begin{center}
  \includegraphics[width=0.6\linewidth]{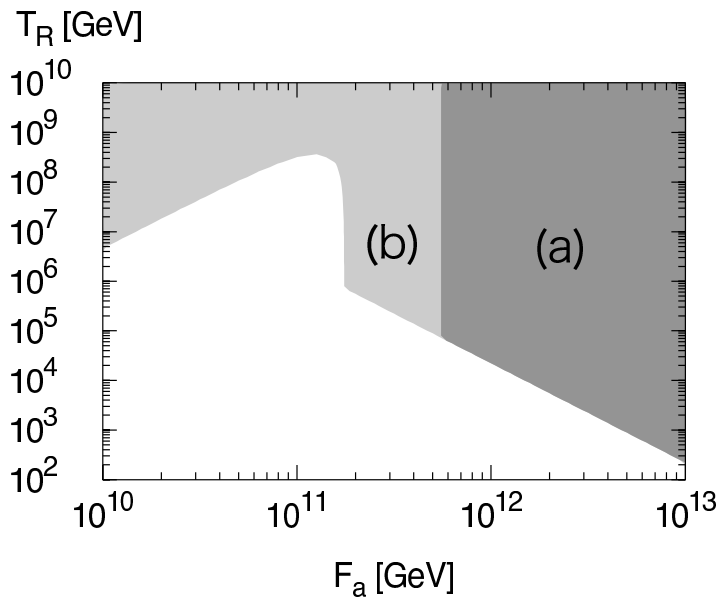}
  \caption{ The shaded regions are exclude (a) for $m_s=10$~eV and 
  	(b) for $m_s=1$~keV. }
  \label{fig:light}
 \end{center}
\end{figure}

In SUSY axion model, the saxion oscillation also contributes to the present dark matter density
as given in Eqs.~(\ref{rhosTos}) and (\ref{rhosTR}).
We should ensure that this contribution does not exceeds the present matter density of the universe.
From Eq.~(\ref{rhosTos}), $\rho_s/s$ is bounded as
\begin{equation}
	\left ( \frac{\rho_s}{s} \right )^{\rm (C)} \lesssim 1.5\times 10^{-11}~{\rm GeV}
	\left ( \frac{m_s}{10~{\rm eV}} \right )^{1/2}
	\left ( \frac{F_a}{10^{11}~{\rm GeV}} \right )^2
	\left ( \frac{s_i}{F_a} \right )^2.
\end{equation}
Thus for $F_a \lesssim 10^{11}$~GeV, the saxion abundance from coherent oscillation is
smaller than that of the dark matter.
However, we should be aware that thermally produced saxion and axino abundances are comparable
to that of the gravitino for $T_R \gtrsim 10^9{\rm GeV}(F_a/10^{11}{\rm GeV})^2$ if 
$m_s = m_{\tilde a} = m_{3/2}$,
since both are thermalized in the early universe.
Thus constraint on the gravitino mass may become more stringent by a factor of two, 
if thermal leptogenesis is assumed to work.
In Fig.~\ref{fig:light}, we show allowed parameter regions on $F_a$-$T_R$ plane
for $m_s=10$~eV (indicated by (a)) and for $m_s=1$~keV (indicated by (b))
with an assumption $s_i =F_a$.
It can be seen that for $m_s \lesssim 10$~eV, thermal leptogenesis is still possible
as long as the PQ scale satisfies $F_a \lesssim 5\times 10^{11}$~GeV.


\section{Conclusions and discussion}   
\label{sec:conclusion}

We have investigated the cosmological constraints on supersymmetric
axion models which are motivated from particle physics point of view.
It is found that the presence of the saxion and the axino makes it
rather difficult to construct a viable cosmological scenario which is
free from any contradiction with observations.  In particular, in almost
all range of the gravitino mass (which is assumed to be the same order as
the saxion mass), the strict upper bound on the reheating temperature is
imposed, and it is more stringent than the bound from the usual
gravitino problem [Figs.~\ref{fig:msTR_10}-\ref{fig:msTR_14}].  The
constraint depends on whether the main decay mode of the saxion is into
axions or not.
It should be noted that
although the axino constraint is stringent for relatively large $m_s$
as can be seen from our results,
it may be significantly relaxed if the axino is much lighter than the  
gravitino.
  The axion is a good candidate for the cold dark matter,
although the axino or gravitino dark matter is also viable for some
parameter regions.

The obtained stringent bound on the reheating temperature has some
implications on the baryogenesis scenarios.  As is well known, the
standard thermal leptogenesis scenario using right-handed neutrino
\cite{Fukugita:1986hr} is incompatible with the gravitino problem except
for $m_{3/2} \lesssim 10$~eV or $m_{3/2} \gtrsim 10$~TeV.  Although the
gravitino mass around $m_{3/2}\sim10$~GeV may also be compatible with
thermal leptogenesis, in this region the NLSP decay into gravitino may
cause another difficulty.  The presence of saxion makes this situation
worse, and hence the standard thermal leptogenesis does not seem to work
in supersymmetric axion models. 
The Affleck-Dine baryogenesis scenario \cite{Affleck:1984fy} can work well
even for such a low-reheating temperature \cite{Kawasaki:2007yy},
except for the case of gauge-mediated SUSY breaking models for small $m_{3/2}$, 
where the Affleck-Dine mechanism may suffer from Q-ball formation 
\cite{Kasuya:2001hg}.
On the other hand, for an ultra-light gravitino mass $m_{3/2} \lesssim 10$~eV,
thermal leptogenesis is still possible.
Here it should be noticed that these constraints from the saxion
strongly depend on the initial amplitude of the saxion $s_i$ and we have
assumed $s_i$ is roughly given by the PQ scale $F_a$.  A concrete
example which gives such initial amplitude is given in
Sec.~\ref{sec:dynamics}.  Perhaps this is the smallest value expected
from naturalness, and hence our bounds presented in this paper should be
regarded as conservative, which means our constraints cannot be relaxed
without significant changes in the cosmological scenario, such as
additional late-time entropy production.

Finally we comment on the detectable signature of the SUSY axion models.
If the saxion mass is around a few MeV, its decay into electron-positron
pair and their annihilation may be observed as the 511 keV line from the
Galactic Center. Such line gamma photons are actually
observed~\cite{Knodlseder:2003sv}.  However, in order to explain the
observed flux, huge entropy production that dilute the saxion density is
needed~\cite{Kawasaki:2005xj}.  Also it may be possible that the nature
of the axion sector is determined from collider experiments if the axino
is the LSP and charged particle such as stau is the NLSP
\cite{Brandenburg:2005he}.

 
\ack{ 
We thank Fuminobu Takahashi and Tsutomu Yanagida for helpful discussion and
comments.
K.N. would like to thank the Japan Society for the Promotion of
Science for financial support.  This work was supported in part by the
Grant-in-Aid for Scientific Research from the Ministry of Education,
Science, Sports, and Culture of Japan, No. 18540254 and No 14102004
(M.K.).  This work was also supported in part by JSPS-AF Japan-Finland
Bilateral Core Program (M.K.). }
 

\section*{References}



\end{document}